\documentstyle[12pt,aasms4]{article}   
\received{** 1999} \accepted{** 1999} \journalid{337}{15 January 1989}
\articleid{11}{14} \slugcomment{}

\begin{document}
\normalsize
\title{Physical conditions in shocked regions of Orion from
ground-based observations of H$_2$O}
\vspace{1.0cm}
\author{Jos\'e Cernicharo}
\affil{Instituto de Estructura de la Materia, Dpto. de F{\'{\i}}sica
Molecular, CSIC, Serrano 121, E-28006 Madrid, Spain}
\affil{e-mail: cerni@astro.iem.csic.es}
\affil{and}
\author{Juan R. Pardo}
\affil{California Institute of Technology, MS 320-47,
Pasadena, CA 91125, USA}
\affil{and}
\author{Eduardo Gonz\'alez-Alfonso}
\affil{Instituto de Estructura de la Materia, Dpto. de F{\'{\i}}sica
Molecular, CSIC, Serrano 121, 28006 Madrid, Spain. Also at
Universidad de Alcal\'a de Henares, Dpto. de F\'{\i}sica,
Campus Universitario, E-28871 Alcal\'a de Henares, Spain}
\affil{and}
\author{Eugene Serabyn, Thomas G. Phillips,
Dominic J. Benford, and David Mehringer}
\affil{California Institute of Technology, MS 320-47,
Pasadena, CA 91125, USA}

\begin{abstract}
We present observations of the 5$_{1,5}\rightarrow$4$_{2,2}$ transition of
water vapor {\it at 325.15 GHz} taken with the CSO telescope
towards Orion IRc2. The emission is more extended than that of other molecular
species such as CH$_3$OH. However, it is much less extended than
the emission of water vapor at 183.31 GHz reported by Cernicharo et al 
(1994). A comparison of the line intensities at 325.15 GHz and 183.31 GHz 
puts useful constraints on the density and temperature of the
emitting regions and allows an estimate of H$_2$O abundance,
x(H$_2$O), of $\simeq$10$^{-4}$ in the Plateau and $\simeq$10$^{-6}$-10$^{-5}$ in 
the Ridge.

\end{abstract}


\section{Introduction}
Water vapor is an important molecule for the chemistry of
interstellar and circumstellar clouds.
The 6$_{16}$-5$_{23}$ masing transition of H$_2$O
at 22 GHz, which arises
from levels around 700K, has been used since
its detection by Cheung et al. (1969) to trace high excitation gas
around star forming regions and evolved stars. The size of the
emitting regions at that frequency is typically
of the order of a few milliarcseconds (a few 10$^{13}$ cm).
Hence, no information has been obtained from this line on the
role of H$_2$O at large spatial scales. Other
H$_2$O lines have been detected from ground or airborne based
telescopes like the 3$_{13}$-2$_{20}$ transition at 183 GHz (Waters
et al., 1980; Cernicharo et al. 1990, 1994, 1996;
Gonz\'alez-Alfonso et al. 1994, 1998), the 4$_{14}$-3$_{21}$
transition at 380 GHz (Phillips, Kwan and Huggins 1980), the
10$_{29}$-9$_{3,6}$ transition at 321 GHz (Menten, Melnick and
Phillips 1990a) and the 5$_{15}$-4$_{22}$ transition at 325 GHz
(Menten et al. 1990b).  Also, the 1$_{10}$-2$_{01}$ transition 
of  H$_{2}^{18}$O at 547 GHz has been observed by Zmuidzinas 
et al. (1994). Among these lines only the 3$_{13}$-2$_{20}$
transition at 183 GHz has been used to map the emission of H$_2$O
at very large spatial scale (Cernicharo et al. 1994, hereafter
referred to as Cer94). The map of the Orion molecular cloud
shown in Cer94 is 6 orders
of magnitude larger than the size of the spots detected at 22 GHz
and for the first time an H$_2$O abundance estimate was derived for
the different large scale components of the Orion molecular cloud.

The ISO satellite has provided the opportunity
to observe thermal lines of water in the middle and 
far-infrared (see the reviews by 
van Dishoeck 1997; Cernicharo 1997; and
Cernicharo et al 1997a,1998).
Mapping of the SgrB2 molecular cloud by
Cernicharo et al. (1997b) has definitely shown that water vapor is
an ubiquitous molecule in molecular clouds with an abundance of
10$^{-5}$.
Maps of the emission of several H$_2$O lines
in Orion IRc2 have been obtained
by Cernicharo et al. (1997a, 1998, 1999). Observations of the central
position have been also obtained by
van Dishoeck et al. (1998), Gonz\'alez-Alfonso et al. (1998) and
Harwit et al. (1998). However, 
The ISO observations of H$_2$O have drawbacks. In addition 
to the limited spectral resolution and the
high opacity of the thermal lines of H$_2$O, the poor
angular resolution provided by ISO in the far-infrared prevents any
detailed study of the spatial structure and physical conditions of the
H$_2$O emitting regions.

An important aid in deriving H$_2$O abundances
could come from
the observation of another masing transition of H$_2$O with
similar properties to those of the 183 GHz line.
The 325 GHz line of H$_2$O was
observed by Menten et al (1990b) in the direction of Orion IRc2 and
other molecular clouds. However, no maps were 
obtained. Here we report the 
detection of extended water emission at
325 GHz and show that the H$_2$O abundance is 
$\simeq$ 10$^{-4}$ in the {\bf Plateau}. The present
data show the importance of ground-based observations of H$_2$O in
deriving the abundance of H$_2$O in molecular clouds and in
providing useful contraints on the physical conditions of the
emitting regions.
Our ground-based observations
provide much finer spatial resolution than ISO or SWAS, 
and an estimate of x(H$_2$O) as accurate as that obtained 
from the extremely optically thick H$_2$O lines observed in the submillimeter
and far infrared domains.

\section{Observations}

The observations were performed with the 10.4 m telescope of the
Caltech Submillimeter Observatory at the summit of
Mauna Kea (Hawaii)
on April 1$^{st}$ 1998.
The receiver, a helium-cooled SIS mixer
operating in double-sideband mode (DSB), 
was tuned at the frequency of the 5$_{1,5}\rightarrow$4$_{2,2}$
line of H$_{2}$O (325.152919 GHz). The H$_2$O line was placed in
the upper sideband (USB) to minimize atmospheric noise from the
image sideband (which was at 322.35 GHz).
Lines in the signal
sideband are severely attenuated relative to those in the image sideband, 
due to the atmospheric H$_{2}O$ line. Therefore it was necessary to 
check the sideband origin of a given line. 
The tuning frequency was shifted by 100 MHz to confirm that
the central feature in the spectrum of Orion-IRC2 was the water vapor 
line. The backend consisted of a 1024 channel acousto-optic spectrometer 
covering a bandwidth of 500 MHz ($\Delta$v=1.1 kms$^{-1}$). Figure 1 
shows the observed spectrum which matches very well the
previous observation by Menten
et al. (1990b) except for the line intensity (see below).

The pointing was determined by observing the same line towards the
O-rich evolved star VY CMa and was found to be accurate to 5''.
The weather conditions were very stable during the observations with
an atmospheric pressure and temperature of 620 mb and -1.4$^{o}$C 
respectively. The relative
humidity was measured to be 4-5\%. The
measured opacity from tipping scans at 225 GHz was $\simeq$0.025.
During the same night we performed broadband Fourier Transform
Spectroscopy (FTS) measurements of the atmospheric absorption
with the FTS described in Serabyn and Weisstein (1995).
Model calculations using the multi-layer atmospheric radiative transfer
model ATM (Cernicharo 1985,1988; Pardo 1996) yielded
an estimated precipitable water vapor column above the telescope of
$\sim$200~$\mu$m, which corresponds to a zenith transmission at 325.15
GHz of $\sim$ 60\% (the corresponding value for the image side bande
was $\sim$ 87\%, hence the line intensities for the image sideband
features are overstimated by a factor $\sim$ 2).

The heterodyne H$_2$O data were calibrated
using an absorber at ambient temperature.
The calculated system noise temperature, for the signal sideband,
was $\simeq$2100 K. The Orion spectrum shown in Figure 1 shows that
the lines from the image sideband are weaker than the 325
GHz H$_2$O line, i.e., just the opposite of that occurring in the
spectrum of Menten et al. (1990b). Rather than a variation
of the maser emission
(the H$_2$O line profile in Figure 1 is
identical to that shown in Menten et al. 1990b)
we think that this
difference is due to much better atmospheric transmission during
our observations. We estimate that our intensity scale is correct
to within 20-30\%.
The Orion-IRc2 map
was carried out in position switching mode by using the
on-the-fly procedure with an off position 5' away in azimuth.
The spatial distribution of the 325 GHz emission is shown in Figure
2 together with that of CH$_3$OH (from the image side band)
and the 183 GHz emission from Cer94. Integrated intensity maps for
selected velocity intervals are also shown in Figure 2.

In order to compare the line profiles of the H$_2$O lines at 183 and 325
GHz we reobserved a few positions at 183 GHz with the 30-m IRAM
telescope in January 1999. The weather conditions were also excellent 
with a zenith opacity at this frequency of $\simeq$ 1. The spectra, together
with those obtained in 1994, are shown in Figure 3.

\section{Results and Discussion}

The observed 5$_{15}$-4$_{22}$ line profile towards the center position
looks similar to that of the 3$_{13}$-2$_{20}$ line observed by Cer94
(see spectra in Figure 3). However, the antena temperature of the line
is 20 times weaker and the line profile, although covering
the same velocity range, is shifted towards the red. Taking into
account the different beam sizes of the IRAM-30m telescope at 183 GHz
and the CSO at 325 GHz, and the extension of the emission in the latter 
line, we estimate that the main beam brightness temperature ratio,
${\bf R}$=T$_{MB}$(183)/T$_{MB}$(325) is 10-20 if both lines
were observed with a telescope of $15''$ beam size. On the other hand,
T$_{MB}$(183)$\sim10^3$ K. Both ${\bf R}$ and T$_{MB}$(183) are well
determined and cannot be related to a calibration problem as the 
atmospheric conditions were extremely good during our observations at 
both frequencies. Similar values for {\bf R}, i.e., ${\bf R}=$10--20, are 
also found at other positions in the cloud (see Figure 2) and represent a 
real difference in the brightness temperatures of the two lines.
Only at position $\Delta\alpha$=-12'', $\Delta\delta$=48'' the peak
temperature of the 5$_{15}$-4$_{22}$ transition approaches that of the
3$_{13}$-2$_{20}$ (the 5$_{15}$-4$_{22}$ line is, however, 
narrower). The 325 GHz emission at this position presents a local maximum 
clearly visible in the velocity maps (see Figure 2).

Like the 3$_{13}$-2$_{20}$ line, the 5$_{15}$-4$_{22}$ transition
is masing in nature. There are some narrow features at 325 GHz
but with intensities of only a few K, i.e., much weaker than those
reported at 183 GHz by Cer94. These features agree in velocity
with those found at 183 GHz. However, ${\bf R}$ changes drastically from
feature to feature, a fact that reveals the maser nature of the 
emission. Outside the central region the lines are very narrow 
(${\Delta}v\simeq$3-5 kms$^{-1}$).

Some of the 3$_{13}$-2$_{20}$ narrow velocity components have antenna 
temperatures above 2000 K and are probably a few arcseconds in size 
(Cer94, Gonz\'alez-Alfonso 1995).
The observations at this frequency performed in January 1999 clearly
indicate a variation in the intensity of some of these features
with respect to those of Cer94. However, at positions
where the line is dominated by the plateau emission
($\Delta\alpha$=28'', $\Delta\delta$=-16'' and
$\Delta\alpha$=12'', $\Delta\delta$=-16'' for example; see Figure 3)
and outside the central region ($\Delta\alpha$=-12'', $\Delta\delta$=48'';
Figure 3) the line shape and intensity do not show any
significant change between both epochs.

A possible explanation for the 183 GHz and 325 GHz emission being 
spatially extended could be that it arises from many masing point like
sources strongly diluted in the beam. This is ruled out 
by the results of Cer94 where 
even the strong features at 183 GHz (T$_{MB}\simeq$2000-4000 K)
show indication of some spatial extent (see above). The densities needed
to reproduce the observed brightness temperatures 
of the maser spots at 22 GHz would result on a thermal or 
suprathermal 183 GHz line. Consequently, if the 183 GHz emission was 
arising from the same region than that of the 22 GHz line 
very large column 
densities will be needed to reproduce the observed 183 GHz and 325 GHz 
intensities. 
In addition, the weak and extended emission observed at 183 GHz by
Cer94 clearly indicates the presence of water vapour coexistent with
the molecular gas in the Orion molecular ridge.

In order to understand the behavior of the two masing lines
we have modeled the radiative transfer of the rotational levels
of p-H$_2$O for the physical conditions of the Orion molecular cloud.
The radiative transfer method is described in Gonz\'alez-Alfonso \& Cernicharo
(1997), and the model consists of a molecular shell with diameter of
$10^{17}$ cm (size of 15$''$ at 450 pc) which expands at a constant velocity 
of 25 km s$^{-1}$. Collisional rates between water vapor and helium were taken
from {\it Green}, Maluendes \& McLean (1993). The Helium 
abundance was assumed to
be 0.1, and the rates were corrected to take into account the collisions
between H$_{2}$O and H$_{2}$.

We calculated the statistical equilibrium populations of the lowest 45
rotational levels of para-H$_2$O for different temperatures (T$_k$=100, 150,
200 and 300 K), column densities N(p-H$_2$O), and volume densities
(n(H$_2$)=$3\,10^5$, $10^6$ and $3\,10^6$ cm$^{-3}$).
Figure 4 shows the main beam brightness temperatures (T$_{MB}$)
(as observed by a telescope of 15$''$ beam size) for the 183 and 325 GHz
para-water lines (thin and broad lines, respectively), together with
${\bf R}$ (dashed lines). These T$_{MB}$ were computed
from the integrated intensity by assuming that the spectral emission is
Gaussian-shaped.

Inspection of Fig. 4 indicates that the line intensity ratio ${\bf R}$
increases with N(p-H$_2$O) for low column densities (which depend on T$_k$
and n(H$_2$)). Both masers are unsaturated in these conditions, but the 
higher opacity of the 183 GHz transition makes this line more sensitive 
to variations of N(p-H$_2$O). For higher values of N(p-H$_2$O), the 183 
GHz line becomes saturated and the exponential
amplification of the 325 maser line yields a decrease of ${\bf R}$.
Finally, when both the 183 and 325 GHz lines are saturated, ${\bf R}$
approaches a nearly constant value or even decreases below 1 for high 
n(H$_2$) and low T$_k$. The maser at 183 GHz is quenched for these later
conditions, although T$_{MB}$ can still remain above T$_k$ due to the 
suprathermal excitation of the line (see Cer94).

Even for relatively low column densities  
(N(p-H$_2$O)$=2\,10^{17}$ cm$^{-2}$), low temperatures (100 K) and
volume density ($3\,10^5$ cm$^{-3}$), the 183 GHz line has an intensity 
larger than 10 K (see Cer94).
However, the possibility of appreciable amplification for the
5$_{15}$-4$_{22}$ line is much more restricted than for the 183 GHz line,
due to the high energy of the levels involved in the 325 GHz line
($\approx450$ K), and to the higher frequency and Einstein coefficient
of this transition. This fact explains the difference in spatial extent 
between both transitions, so that the the 325 GHz line is spatially 
restricted to the Plateau while the 183 GHz line is in addition detected 
in the Ridge (Cer94).

The water vapor column density that fits the observed log {\bf R}$=1-1.3$
depends strongly on the assumed values of n(H$_2$) and T$_k$. The higher
n(H$_2$) and T$_k$, the lower N(p-H$_2$O) that is needed to obtain an appreciable
amplification of the 325 GHz line. Fig. 4 shows
that {\bf R}$=10-20$ is obtained in different panels for N(p-H$_2$O) ranging
from $3\,10^{17}$ cm$^{-2}$ (n(H$_2$)=$3\,10^6$ cm$^{-3}$, T$_k$=300 K)
to $2\,10^{19}$ cm$^{-2}$ (n(H$_2$)=$3\,10^5$ cm$^{-3}$, T$_k$=100 K).
However, some of these models are not compatible with the observed
intensities. For n(H$_2$)$\ge$$10^6$ cm$^{-3}$ and T$_k$$>$150 K,
we find that {\bf R}$=20$ yields T$_{MB}$(325) in excess of $10^2$ K
and T$_{MB}$(183)$>$$2\,10^3$ K. Both the observed intensities and
{\bf R} are only compatible with more moderate values of
n(H$_2$) and/or T$_k$. The physical reason is that, for high values of
n(H$_2$) and/or T$_k$, the collisional pumping of the 183 GHz line
becomes so efficient that the emission in this line reaches high intensities 
for column densities that provide T$_{MB}$(325) of 50--100 K. Of course,
the Plateau may have regions with very high densities and temperatures
(which will give rise, for example, to the emission at 22 GHz and to the 
narrow spectral features observed at 183 and 325 GHz), but these will
be much smaller than the observed size of the cloud. The widespread 
emission from the Plateau we observe at 183 GHz and 325 GHz is only compatible
with moderate values of n(H$_2$) and T$_k$.
In our models, the 22 GHz line will have intensities 
similar to those already observed in
Orion (Genzel et al, 1981) only for the highest kinetic temperatures and 
volume densities in Figure 4. 
The brightest spots at
22 GHz could be correlated with the narrow features at 183 GHz, and
with the relatively weak lines at 325 GHz.
Brightness temperatures above 10$^6$ K can be obtained for large
column densities, T$_K$$>$150 K and n(H$_2$)$>$10$^6$ cm$^{-3}$.

Lower limits for n(H$_2$) and T$_k$ can be obtained from the observations
of other molecular lines (e.g., Blake et al. 1987), so that
we adopt n(H$_2$)$\ge$$10^6$ cm$^{-3}$ and T$_k$=100--150 K. For these
conditions we obtain N(p-H$_2$O) in the range $2\,10^{18}$--$5\,10^{18}$ 
cm$^{-2}$, and hence N(H$_2$O) in the range $8\,10^{18}$--$2\,10^{19}$
cm$^{-2}$. The water vapor abundance can be derived from the CO data
taken with similar angular resolution (see Cer94).
For the intermediate velocity wind Cer94 derived a CO column density
of 10$^{19}$ cm$^{-2}$. Hence, the x(H$_2$O)/x(CO) abundance ratio in the 
Plateau is around 1, i.e., x(H$_2$O)$\sim$$1-2\,10^{-4}$. 
In the Ridge molecular
cloud the 325 GHz is very weak. Our models and the 183 GHz data
provide an estimate for x(H$_2$O) of a few 10$^{-6}$-10$^{-5}$ (see Cer94)
which is in good agreement with our ISO results (see Cernicharo et al. 1998,
1999).

The comparison of several masing transitions arising in relatively
low energy levels of H$_2$O allows us to constrain the
physical conditions of the different emitting regions.
So far, ground-based
observations of these transitions with large
radio telescopes are the only  means to obtain the spatial
distribution of H$_2$O in interstellar clouds.

\begin{center} {\it Acknowledgements} \end{center}
J. Cernicharo and E. Gonz\'alez-Alfonso acknowledge Spanish DGES for this
research under grants PB96-0883 and ESP98-1351E.
J.R. Pardo gratefully acknowledges
the financial support of the 
{\it Observatoire de Paris-Meudon}, {\it CNES} and  
{\it M\'et\'eo-France}. The CSO is supported by 
NSF contract \# AST-9615025.


\clearpage

\begin{center}
Figure Captions
\end{center}         

\figcaption[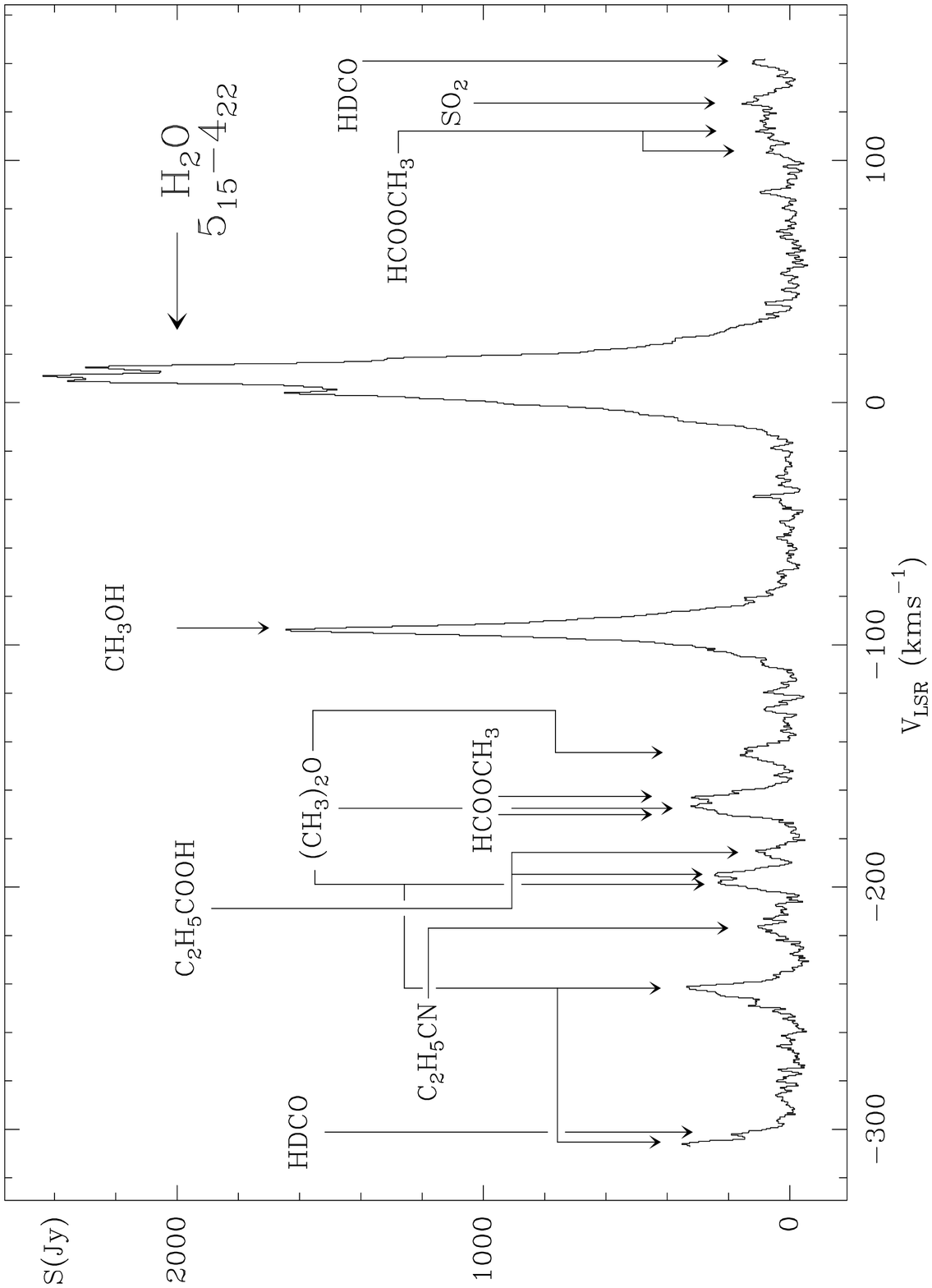]
{Observed spectrum at 325.153 GHz
in the direction of Orion IRc2. All lines,
except the 5$_{15}$-4$_{22}$ transition of H$_2$O and one of the lines 
of methyl formate (at +100 kms$^{-1}$), are from the image sideband.
The Y-axis units are in Jy for the signal side band.
The intensities for the lines in the image side band 
are severely overestimated. 
The 5$_{15}$-4$_{22}$ line of H$_2^{18}O$ at 322.464 GHz (image side band at
V$_{LSR}$=100 kms$^{-1}$) is overlapping with a
line of CH$_3$COOH.}

\figcaption[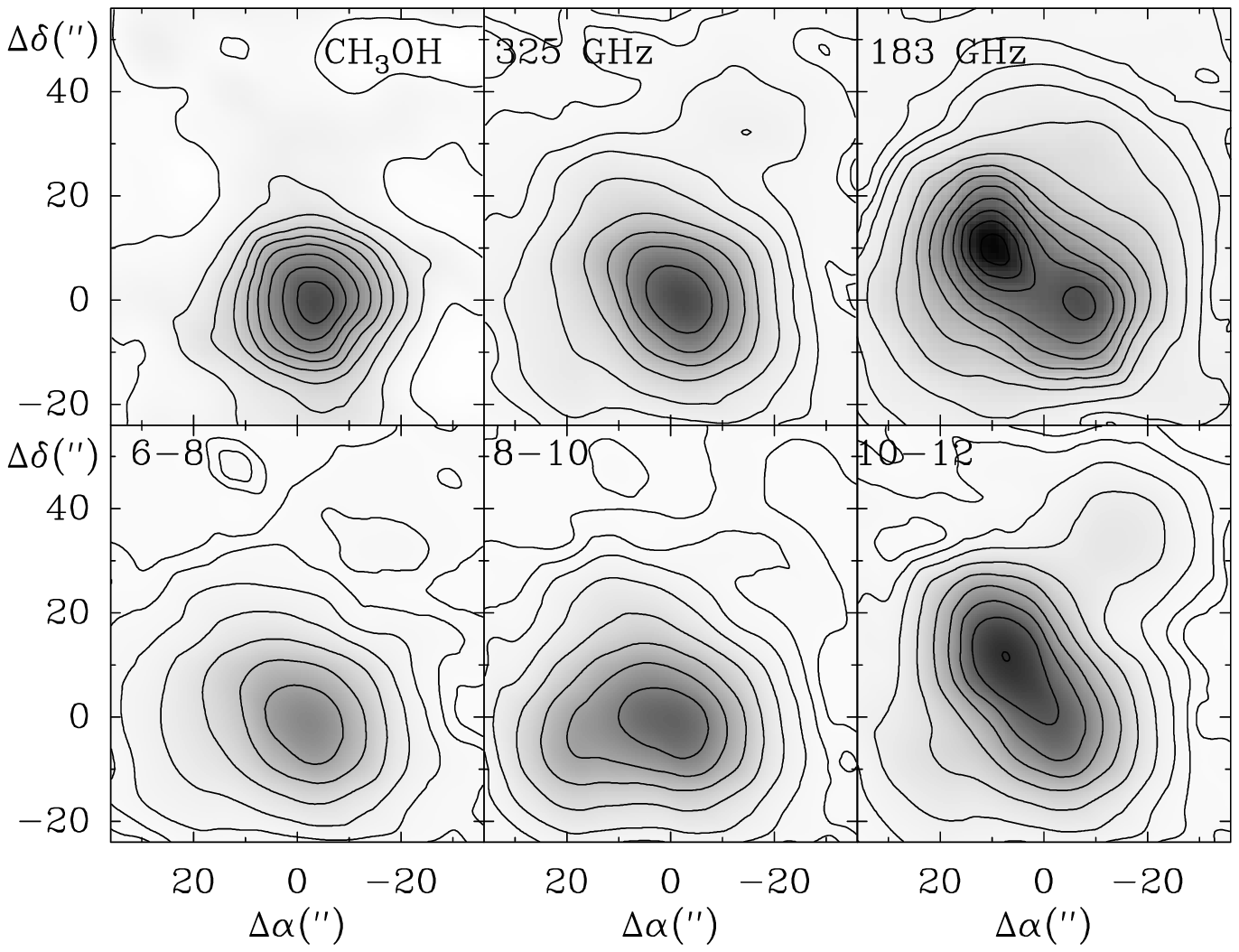]
{Integrated line intensity between -20 and 40 kms$^{-1}$
for CH$_3$OH (top left), H$_2$O at 325 GHz (5$_{15}$-4$_{22}$)
(top middle), and H$_2$O at 183 GHz (3$_{13}$-2$_{20}$) (top
right left; from Cer94). The CH$_3$OH data are uncalibrated as
the line is in the image side band.
Black contours for H$_2$O at 325 GHz are 0,
10, 50, 100, and 200  to 1200 by 200 K$\;$kms$^{-1}$, and for H$_2$O at 183 
GHz they are 100, 300, 600, 1000 and 2000 to 18000 by 2000 K$\;$kms$^{-1}$.
The other panels show the integrated line intensity of H$_2$O at 325 GHz for
selected velocity intervals (6-8, 8-10 and 10-12 kms$^{-1}$).}

\figcaption[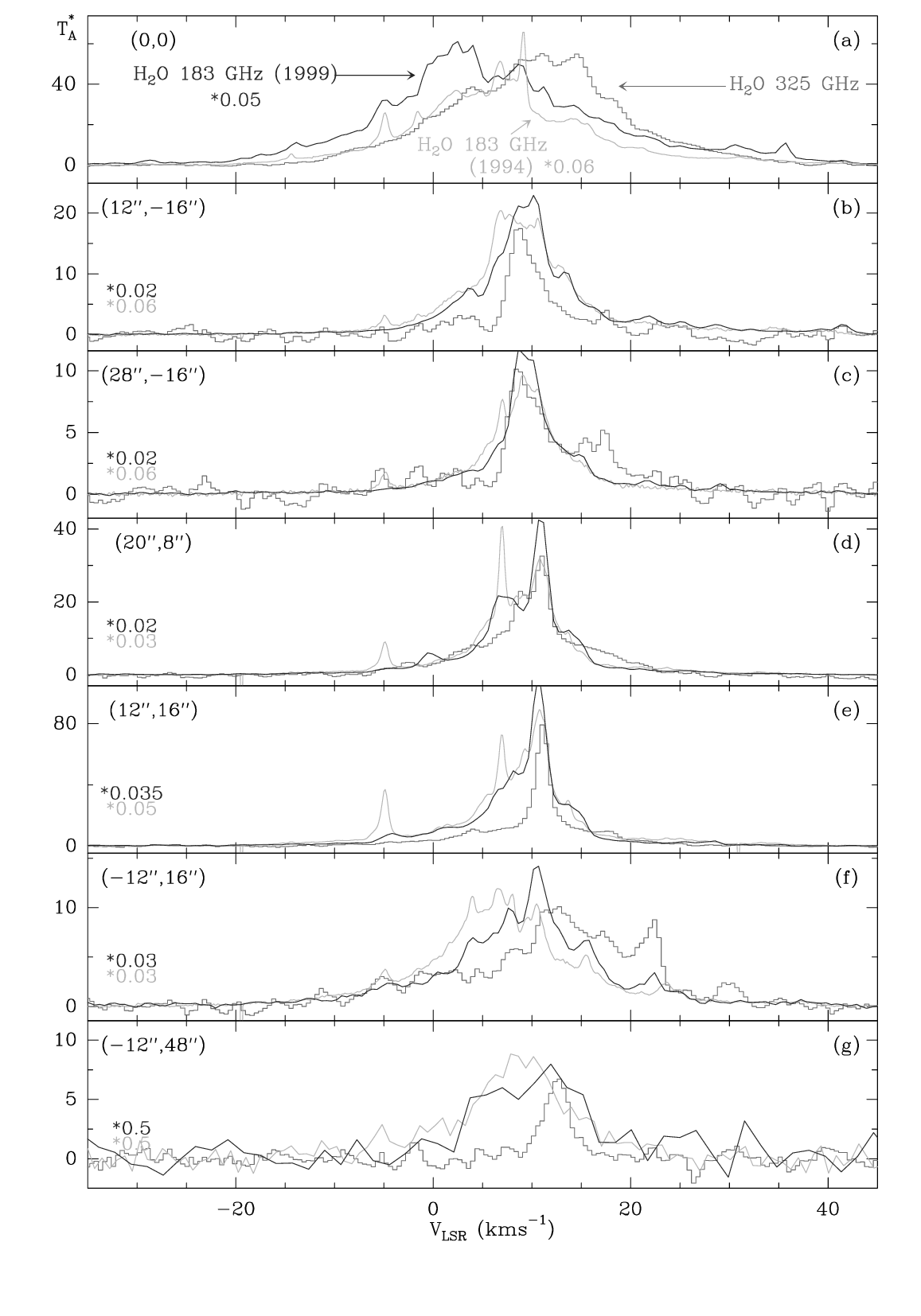]
{Line profiles of the H$_2$O (5$_{15}$-4$_{22}$)
transition at selected positions (thick histogram). The corresponding 
rescaled data for the (3$_{13}$-2$_{20}$) line at 183 GHz as observed
in 1999 are shown as continuous black lines, while those obtained in 1994
are plotted as solid thin lines. The scaling factors for the 183 GHz
observations are indicated in each panel.}

\figcaption[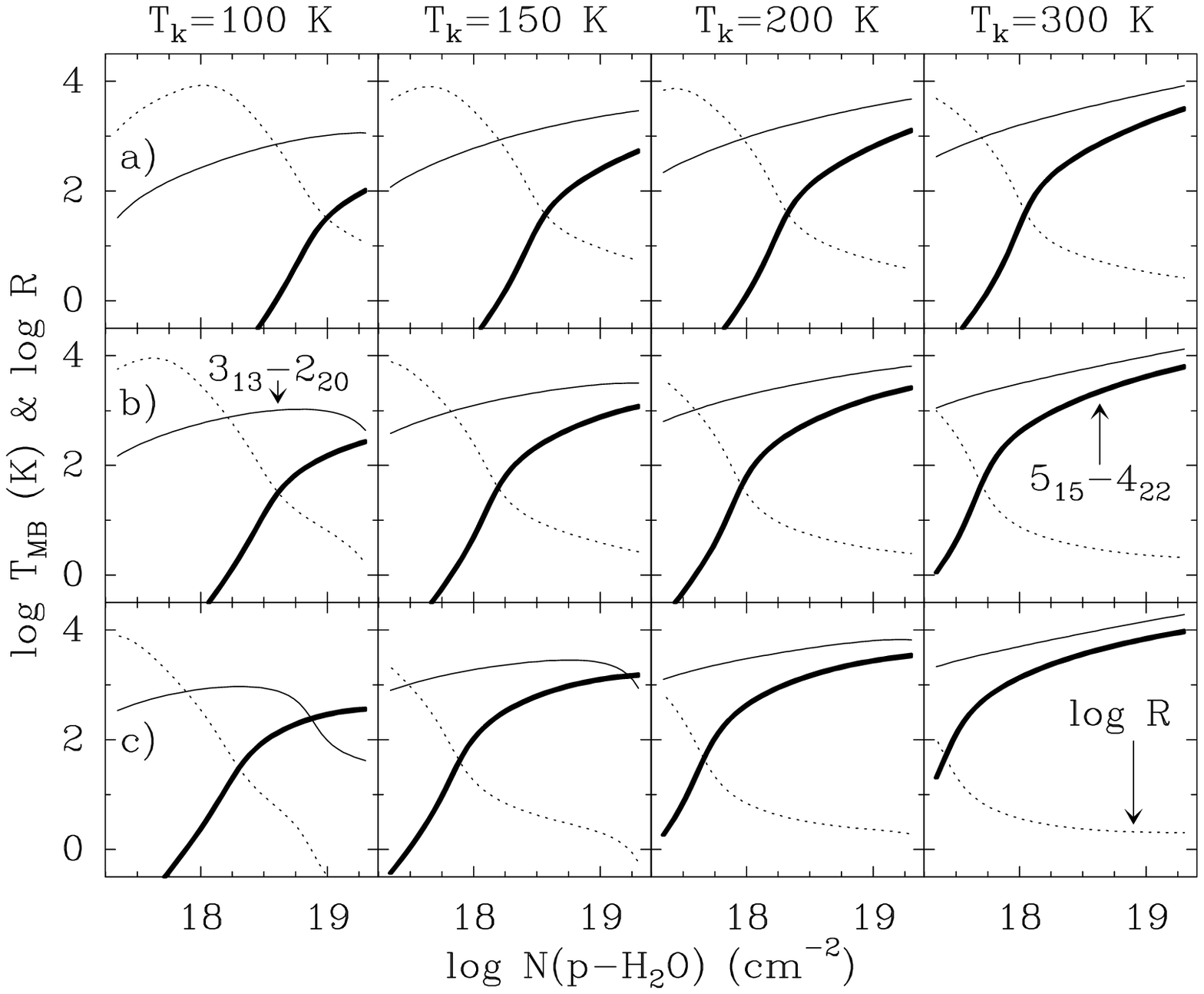]
{Results for the brightness main beam temperature of the
3$_{13}$-2$_{20}$ (183.31 GHz), and the
5$_{15}$-4$_{22}$ (325.15 GHz) vs. the p-H$_2$O column density. 
Each column corresponds to a different kinetic temperature
(100, 150, 200 and 300 K from left to right), while
each row corresponds to a different molecular hydrogen density
(n(H$_2$)=3 10$^{5}$, 10$^{6}$,
and 3 10$^{6}$ cm$^{-3}$ -rows {\bf a}, {\bf b} and
{\bf c}).}

\clearpage

\begin{figure}[h]
\plotone{fig1.ps}
\end{figure}

\clearpage                                        

\begin{figure}[h]
\plotone{fig2.ps}
\end{figure}

\clearpage

\begin{figure}[h]
\plotone{fig3.ps}
\end{figure}

\clearpage

\begin{figure}[h]
\plotone{fig4.ps}
\end{figure}

\end{document}